\documentclass[aps,prl,reprint,groupedaddress,showpacs]{revtex4-1}

\usepackage{graphicx}

\begin{document}


\title{Neutron skin thickness of heavy nuclei with $\alpha$-particle
  correlations \\ and the slope of the nuclear symmetry energy}



\author{S. Typel}
\affiliation{GSI Helmholtzzentrum f\"{u}r Schwerionenforschung GmbH,
Planckstr. 1, D-64291 Darmstadt, Germany}


\date{\today}

\begin{abstract}
The formation of $\alpha$-particle clusters on the surface of
heavy nuclei is described in a generalized relativistic mean-field model 
with explicit cluster degrees of freedom. The effects on the size of the 
neutron skin of Sn nuclei and ${}^{208}$Pb are investigated as a function of the
mass number and the isospin-dependent part of the effective
interaction, respectively. The correlation of
the neutron skin thickness with the difference of the neutron
and proton numbers and with the slope of the nuclear symmetry
energy is modified as compared to the mean-field
calculation without $\alpha$-cluster correlations.
\end{abstract}

\pacs{21.10.Gv,21.60.Gx,21.60.Jz,21.65.Ef}

\maketitle


Correlations are an essential feature in interacting many-body
systems, which can have a strong impact on particular observables. 
In dilute nuclear matter, the strong interaction leads to the appearance of
clusters, i.e.\ correlated states of nucleons,
below the nuclear saturation density $n_{\rm sat} \approx
0.15$~fm$^{-3}$. Such conditions are found
in the debris of heavy-ion collisions when the hot compressed
baryonic matter expands, cools and
fragments of different sizes emerge, see, e.g.\ Refs.\ \cite{Zha12,Qin12,Lin14}. 
In the post-bounce evolution of core-collapse supernovae 
large abundancies of light clusters might affect the neutrino absorption
and heating of the low-density matter behind the shock front 
\cite{Sum08,Arc08,Fur13}.
On the surface of nuclei, the formation of clusters is a
prerequisite for cluster radioactivity \cite{Del10} and in particular the
$\alpha$-decay of heavy nuclei \cite{Gam28}. Here, the tunneling
through the Coulomb barrier is well understood but the preformation of
the $\alpha$-particle is a challenge of theoretical model descriptions.
Clustering phenomena 
are expected to affect the density dependence of the symmetry energy of nuclear matter 
\cite{Typ14} and
the structure of nuclei, in particular skin and halo
phenomena, see, e.g., the review article \cite{Hor12}. The natural emergence
of clusters is still a difficult task in many nuclear structure models.

In this work, the effect of $\alpha$ clustering on the neutron skin
thickness of heavy nuclei is investigated. This quantity is defined as the
difference
$ r_{\rm skin} = r_{n} - r_{p}$
of the root-mean-square (rms) radii of neutrons, $r_{n}$, and protons, $r_{p}$.
A strong correlation of the neutron skin thickness with the density dependence of 
the neutron-matter equation of state \cite{Bro00,Typ01} and that of
the symmetry energy of nuclear matter, see e.g.\ Refs.\
\cite{Furnstahl:2001un,Vinas:2013hua}, 
was found in mean-field calculations.
At present, the density dependence of the symmetry energy is intensively studied
in theory and experiment using different approaches, see
the articles in the topical issue \cite{EPJA}.
It can be quantified with the so-called slope 
coefficient $L$ that
appears in the expansion of the energy per baryon in nuclear matter,
see, e.g., Ref.\ \cite{Typ14}.
A precise knowledge of the
relation between $r_{\rm skin}$ and $L$ and consequently the density dependence
of the symmetry energy is essential 
for predicting the structure of neutron stars,
in particular their radii \cite{Ste13}. Hence, there are a large number of 
experimental attempts in recent years in order to determine either
the neutron skin thickness $r_{\rm skin}$
directly, e.g.\ by parity violation in electron scattering
on Pb nuclei in the PREX experiment \cite{Abr12a,Abr12b}, 
or the slope coefficient $L$  by indirect methods, see, e.g., Refs.\
\cite{Tsang:2012se,Lattimer:2012xj,Vinas:2013hua}
and references therein.
Until now, the quantitative correlation between $r_{\rm skin}$ and $L$ relies on
the description of nuclei in self-consistent mean-field approaches
\cite{Ben03} such as nonrelativistic
Skyrme Hartree-Fock and relativistic mean-field (RMF) calculations.
These models are based on the picture of independent nucleonic
quasi-particles. They do not
consider residual cluster correlations beyond pairing in the most simple applications.

In the external region of a
nucleus low-density nuclear matter properties are tested. 
Such dilute matter at finite temperature 
is described by models for the equation of state
that are designated for astrophysical applications \cite{Hem11}.
Many-body correlations have to be taken into account in order to describe correctly
the thermodynamic properties and chemical composition 
of the system, most notably the formation
of light clusters, such as deuterons or $\alpha$-particles. 
Therefore, similar effects can be anticipated in the vicinity of the nuclear
surface.

In Refs.\ \cite{Typ10,Vos12,Typ14} an extended RMF
model with density
dependent couplings was developed that treats few-body correlations as
explicit degrees of freedom. The formation and dissolution of
clusters are a result of medium dependent mass shifts, which are taken
from a quantumstatistical approach to describe clusters in dilute matter. These shifts
originate mainly from the action of the Pauli exclusion principle that
prohibits the formation of few-body bound and resonant states with increasing
density of the medium. The model was applied to the description of 
finite nuclei in warm matter
applying fully self-consistent calculations in an
extended relativistic Thomas-Fermi (RTF) approximation within spherical
Wigner-Seitz cells \cite{Typ13a}. It was found that a heavy nucleus is formed in the
center of the cell, which is surrounded by a low-density gas of nucleons and light
clusters. A particular observation was the enhanced probability of
finding clusters on the nuclear surface, 
see Fig.~11 in Ref.\ \cite{Typ13a}.
This behavior is caused by an attractive pocket in the effective
cluster potentials at the nuclear surface due to
a finite range of the interaction.
The attractive
scalar potential $S_{i}$ extends further out than the repulsive vector
potential $V_{i}$ of a cluster $i$.
Typical values of 1-2~fm and 10-20~MeV for the width and the depth,
respectively, of the potential pocket are obtained in the present calculations.
In contrast, the appearance of clusters inside the heavy nucleus is strongly
suppressed because of the large positive mass shift in the scalar
potential.

\begin{figure}
\begin{center}
\includegraphics[width=0.9\linewidth]{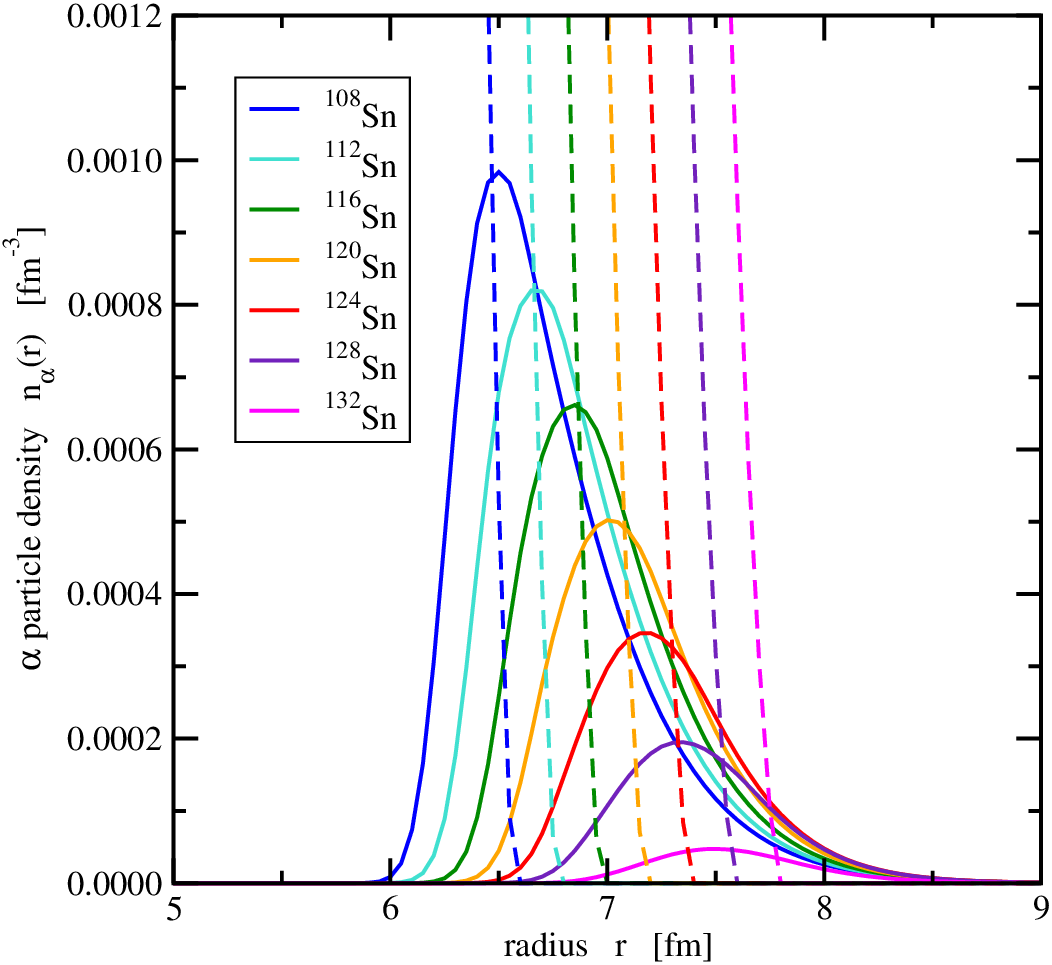}
\end{center}
\caption{\label{fig:alpha_distr_sn}%
(Color online) Radial density distribution of $\alpha$-particles (full
lines) and neutrons (dashed lines) for a selected set of isotopes of
the Sn chain from ${}^{108}$Sn (leftmost) to ${}^{132}$Sn (rightmost).}
\end{figure}

A similar 
approach can be used to describe heavy nuclei in the vacuum at
zero temperature in order to study the significance of few-body correlations at
the nuclear surface. However, a few modifications 
have to be taken into account. The $\alpha$-particle with the highest binding energy of
the light clusters emerges as the only relevant correlation. 
In contrast to nucleons, which are fermions and
can be treated in the Thomas-Fermi approximation,
$\alpha$-particles are bosons.
They populate only the ground state wave function,
which has to be determined explicitly. This 'condensation' is one
foundation of the very successful THSR description of dilute excited
nuclei \cite{Zho13}, e.g. the Hoyle state in ${}^{12}$C, where the
many-body wave fuction is constructed from $\alpha$-particles
occupying the same quantum state.
In the present calculation, a
WKB approximation is used to obtain the $\alpha$-particle wave
function self-consistently with the nucleon distributions.
The resulting density distribution of ${}^{4}$He has a
maximum at the position of the pocket in the effective potential (see below).
The amount of $\alpha$-clustering is determined such that the
effective position-dependent $\alpha$ energy $E_{\alpha}(\vec{r}) = m_{\alpha} +
V_{\alpha}(\vec{r}) - S_{\alpha}(\vec{r})$ does not exceed the
$\alpha$-particle chemical potential $\mu_{\alpha} = 2 \mu_{n} + 2
\mu_{p}$ (including rest masses). The latter is
given by the neutron and proton chemical potentials $\mu_{n}$ and
$\mu_{p}$ that are found from the extended RTF description
of the nucleon distributions. The number of $\alpha$-particles is a
result of the self-consistent solution of the coupled equations with
nucleon, $\alpha$-particles and meson fields as degrees of freedom.
The radial distribution $n_{\alpha}(r)$ is given by the 
modulus square $n_{\alpha} = \left| \psi_{\alpha}\right|^{2}$ of the
$\alpha$-particle wave function $\psi_{\alpha}(\vec{r})$ and the
absolute number is found in a variational calculation.

\begin{figure}
\begin{center}
\includegraphics[width=0.9\linewidth]{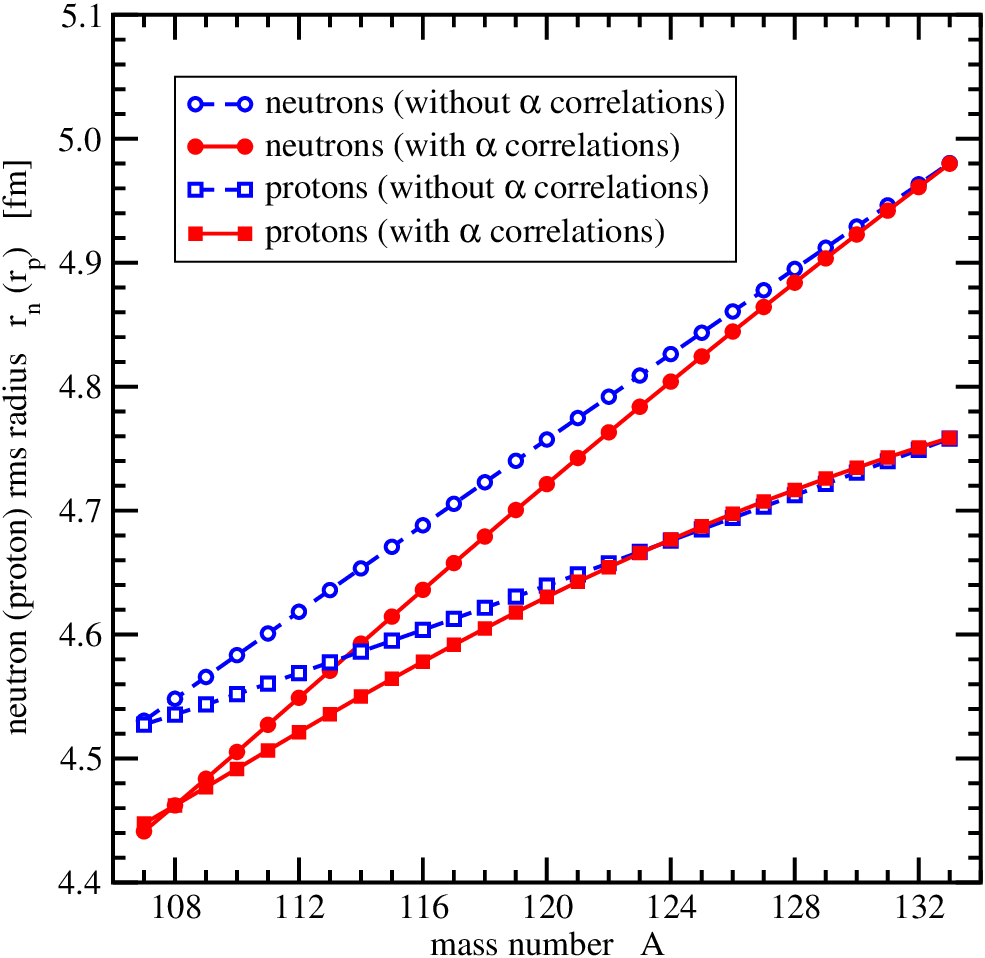}
\end{center}
\caption{\label{fig:rms_sn}%
(Color online) Dependence of the rms radii of neutrons
(circles) and protons (squares) on the mass number
$A$ of Sn nuclei in the extended RTF calculation with the
modified DD2 parametrization. Full red (open blue) symbols denote the results
with (without) $\alpha$-particle correlations.}
\end{figure}

The density-dependent DD2
parametrization was introduced in the extended RMF model of Ref.\
\cite{Typ10}. 
It was obtained by fitting the parameters 
to properties of finite nuclei using the usual mean-field
Hartree approximation. It can be directly applied to the description
of homogeneous matter with clusters as in Refs.\ \cite{Typ10,Vos12}, however, the
calculation of nuclear properties in the extended RTF approximation
will give slightly different results for energies and radii. In order
to compensate, at least partly, for these differences, in the present calculations 
the mass of the $\sigma$ meson was increased from the original value
$m_{\sigma}^{\rm (orig)}$ 
to $m_{\sigma}^{\rm (mod)}=577.9$~MeV and the $\sigma$ meson coupling
$\Gamma_{\sigma}$ was multiplied by the factor $m_{\sigma}^{\rm
  (mod)}/m_{\sigma}^{\rm (orig)}$. This rescaling does not affect the
results for uniform matter but it improves the description of finite nuclei.
Although the extended TF calculations will give smaller
neutron skin thicknesses than the full Hartree calculations (see
below), the general trends due to the
$\alpha$-particle correlations can be studied in such an approach.

The radial distributions of $\alpha$-particles for seven isotopes of
the Sn chain is shown in Fig.~\ref{fig:alpha_distr_sn} by full
lines. The corresponding distributions for neutrons are indicated by
dashed lines with a very steep decrease with increasing radius.
It is obvious that the $\alpha$-particle densities
are much smallar than those of the nucleons.
The position of the maximum in the $\alpha$-particle
density $n_{\alpha}(r)$ moves to larger radii in accordance with the
extension of the neutron distribution when the neutron excess of the
Sn nuclei increases. At the same time, the height of the maximum
decreases significantly such that the total amount of
$\alpha$-particles at the surface becomes smaller.

\begin{figure}
\begin{center}
\includegraphics[width=0.9\linewidth]{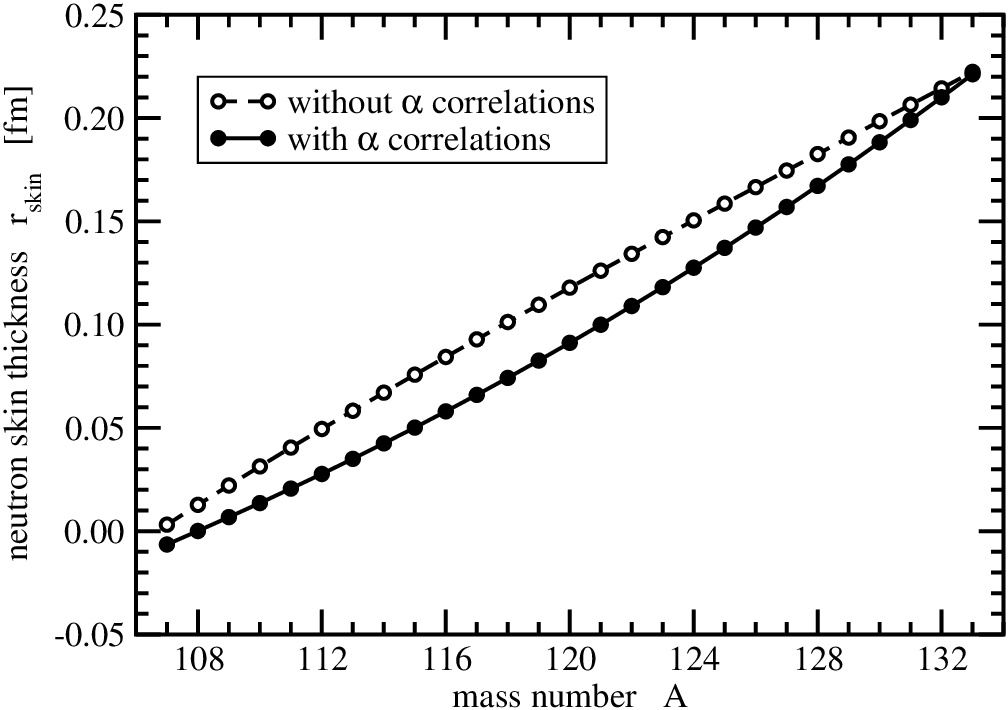}
\end{center}
\caption{\label{fig:skin_sn}%
Dependence of the neutron skin thickness
on the mass number $A$ of Sn nuclei in the extended 
RTF calculation with the
rescaled DD2 parametrization. Full (open) symbols denote the results
with (without) $\alpha$-particle correlations.}
\end{figure}

In Fig.~\ref{fig:rms_sn} the evolution of the neutron and proton
rms radii in the chain of Sn isopotes is depicted when
the mass number $A$ increases. At mass numbers $A\approx 107$ the 
neutron and proton distributions of a nucleus have almost identical 
rms radii without forming a neutron skin. At even lower mass numbers a
proton skin develops with a size that could also be affected by
$\alpha$-clustering. However, for lower $A$ the $\alpha$-particle
becomes unbound in the present model and the $\alpha$-particle dripline is crossed.
With increasing neutron
number, the neutron rms radius rises stronger than the proton rms
radius and a neutron skin appears. 
With $\alpha$-particle correlations, 
however, the rms radii are smaller for a given
nucleus than in the model without $\alpha$ correlations. 
This is a consequence of the larger diffuseness of the total neutron and
proton density distribution. It requires smaller rms radii to keep
the total number of neutrons and protons for a given nucleus constant.
For $A \approx 133$ the differences between the rms radii in the model
calculations without and with $\alpha$-particles practically vanish.

\begin{figure}
\begin{center}
\includegraphics[width=0.9\linewidth]{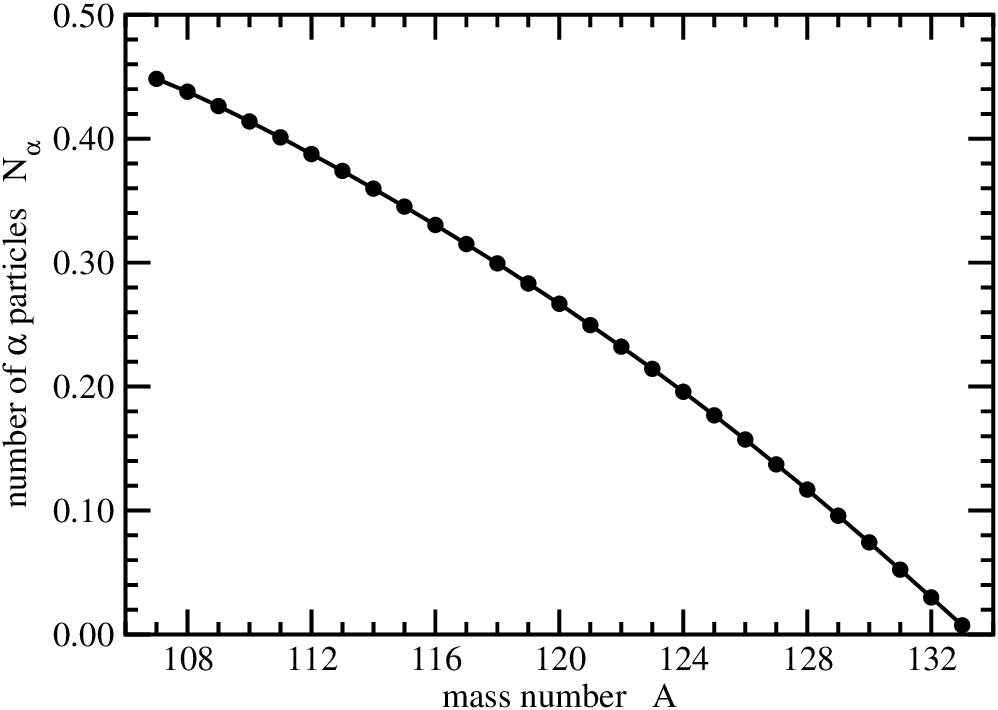}
\end{center}
\caption{\label{fig:alpha_sn}%
Dependence of the number of $\alpha$-particles $N_{\alpha}$
on the mass number $A$ of Sn nuclei in the extended RTF
calculation with the rescaled DD2 parametrization.}
\end{figure}

The dependence of the resulting neutron skin thicknesses $r_{\rm
  skin}$  on $A$ for the
same chain of Sn nuclei in shown in Fig.~\ref{fig:skin_sn}. Without
$\alpha$-correlations,
$r_{\rm skin}$ increases almost linearly with the mass number $A$. 
The values of the present calculation and their mass number
dependence are comparable to the results of the
Hartree-Fock-Boguliubov calculations with the models BSk24, BSk25, and
BSk26 in Ref.\ \cite{Goriely:2013xba}.
However, the consideration of $\alpha$-cluster correlations leads to a
substantial reduction of the neutron skin, in particular in the middle
of the chain. This can be well understood because the appearance of
$\alpha$-particles on the nuclear surface pushes the abundancies
of neutrons and protons (including those bound in clusters) 
towards a more symmetric distribution.
For small $A$ with almost the same neutron and proton
numbers in the nucleus, there is no effect and no neutron skin
develops. For a large neutron excess, $\alpha$-particles cannot be
formed efficiently in the neutron-rich low-density matter on the
nuclear surface and the effect vanishes again.

The effects observed in Figs.~\ref{fig:rms_sn} and \ref{fig:skin_sn}
correlate with the amount of $\alpha$-particles that appear on the
surface of the nucleus.  
The effective number $N_{\alpha}$ of $\alpha$-particles in
the nuclei of the Sn chain is illustrated in
Fig.~\ref{fig:alpha_sn}. Since the present approach is based on a
statistical description, $N_{\alpha}$ is not an integer number.
For small $A$ the effective $\alpha$-particle number is largest.
With decreasing $A$, the binding energy of an $\alpha$ cluster reduces 
and finally the $\alpha$-particle drip
line will be reached, indicating the possibility of $\alpha$-decay.
By increasing the mass number $A$ in the chain of Sn nuclei, the effective
number of $\alpha$-particles at the nuclear surface decreases continuously until it finally
vanishes for large $A$. Here, a sizeable neutron skin develops but the 
four-nucleon correlations have no effect on its size since $\alpha$-particles
do not form in a significant amount in such a neutron-rich environment.

The formation of $\alpha$-particle correlations at the nuclear surface
will modify the universal relation between the neutron skin thickness 
$r_{\rm skin}$ and the symmetry energy slope coefficient $L$
that was established in mean-field descriptions of nuclei and nuclear matter.
The size of the neutron skin of heavy nuclei is strongly affected by
the density dependence of the symmetry energy that reflects the
isospin dependence of the nuclear interaction. 
In RMF models the isovector $\rho$ meson usually
represents the only contribution to the isospin
dependence of the interaction. Earlier versions with nonlinear meson
self-interactions
considered only a single parameter, the $\rho$ meson coupling strength
$\Gamma_{\rho}$.
In the RMF approach with
density dependent meson-nucleon couplings and
parametrizations such as TW99 \cite{Typ99}, DD2 \cite{Typ10}, 
\dots, the $\rho$ meson coupling
\begin{equation}
 \Gamma_{\rho}(n) = \Gamma_{\rho}(n_{\rm ref}) \exp \left[ - a_{\rho}
 \left( \frac{n}{n_{\rm ref}} - 1\right) \right]
\end{equation}
depends on the total baryon density $n$ with three parameters: the coupling 
$\Gamma_{\rho}(n_{\rm  ref})$ at a reference density $n_{\rm ref}$
(usually taken as the saturation density $n_{\rm sat}$)
and a parameter $a_{\rho}$ that regulates the strength of the density
dependence.
A variation of $\Gamma_{\rho}(n_{\rm ref})$ and $a_{\rho}$ modifies
the symmetry energy at saturation density $J$ and in particular the
density dependence characterized by the slope coefficient $L$. 

\begin{table}[t] 
\caption{\label{tab:dd2var}%
Isovector parameters for the variations of the DD2 parametrization of
the RMF model with density dependent meson-nucleon couplings.}
\begin{ruledtabular}
\begin{tabular}{lcccc}
parametrization & symmetry & slope & $\rho$-meson & $\rho$-meson \\ 
 & energy & coefficient  & coupling & parameter \\
 & $J$ [MeV] & $L$ [MeV] & $\Gamma_{\rho}(n_{\rm ref})$ &  $a_{\rho}$ \\
\hline
 DD2$^{+++}$ & 35.34  & 100.00 & 4.109251 &   0.063577 \\ 
 DD2$^{++}$  & 34.12  &  85.00 & 3.966652 &   0.193151 \\
 DD2$^{+}$   & 32.98  &  70.00 & 3.806504 &   0.342181 \\
 DD2    & 31.67  &  55.04 & 3.626940 &   0.518903 \\
 DD2$^{-}$   & 30.09  &  40.00 & 3.398486 &   0.742082 \\
 DD2$^{--}$  & 28.22  &  25.00 & 3.105994 &   1.053251 \\
\end{tabular}
\end{ruledtabular}
\end{table}

In order to study the correlation between the neutron skin thickness 
and the slope coefficient, variations of the original DD2
parametrization were created by fixing $L$ to
particular values and refitting $J$ to properties of finite nuclei.
In this process the isoscalar part of the effective interaction, i.e.\
the $\sigma$ and $\omega$ meson couplings and their density dependence
were not touched. In Table \ref{tab:dd2var}, the parameters of these
new effective interactions are given. Parametrization with $L$ values
larger than the original DD2 value are denoted by DD2$^{+++}$,
DD2$^{++}$, and DD2$^{+}$. In contrast, the parametrizations 
DD2$^{-}$ and DD2$^{--}$ have smaller values for $L$. The correlation 
of $J$ and $L$ is obvious. Larger slope coefficients are
accompanied by larger symmetry energies at saturation. It has to be
mentioned that the quality of the description of finite nuclei 
deteriorates when $L$ deviates strongly from the value of the
original DD2 parametrization, but the variation covers the range of
typical mean-field model calculations.

\begin{figure}
\begin{center}
\includegraphics[width=0.9\linewidth]{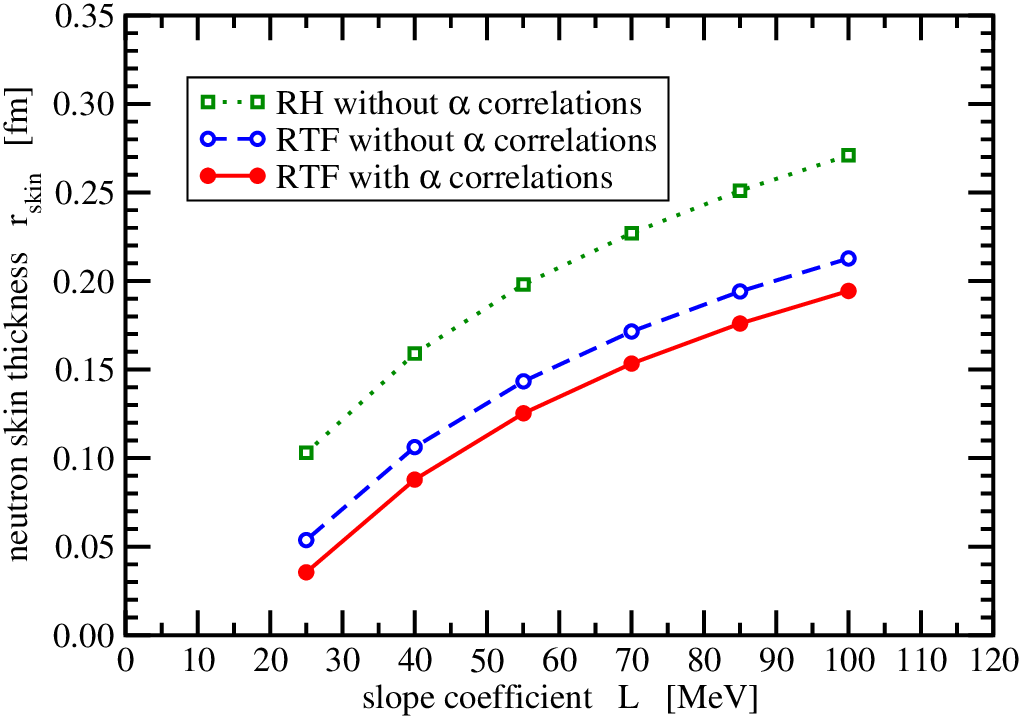}
\end{center}
\caption{\label{fig:skin_pb}%
(Color online) Dependence of the neutron skin thickness of Pb nuclei
on the slope parameter $L$. Squares denote the results of the RMF
calculation with the original DD2 parametrization in Hartree
approximation and full (open) circles are those of the relativistic TF model 
with the rescaled DD2 parametrization
with (without) $\alpha$-particle correlations.}
\end{figure}

The correlation between the neutron skin thickness $r_{\rm skin}$ of
the ${}^{208}$Pb lead nucleus
and the slope coefficient $L$ of the 
nuclear symmetry energy is depicted in Fig.~\ref{fig:skin_pb}
for the six different parametrizations of Table \ref{tab:dd2var}.
A distinct correlation between $r_{\rm skin}$ and $L$ is observed that
is well known from previous mean-field calculations.
The green open squares show the correlation in the original mean-field
Hartree calculation that was used to fit the parameters of the 
interactions. The neutron skin thickness rises with increasing $L$.
Since the parameter sets with $L$ values departing from that of the
original DD2 parametrization are not optimal fits to all considered
properties of finite nuclei and the isoscalar part of the effective
interaction is not modified, a curvature of the correlation is
found in contrast to the almost linear correlation that is observed
for models with best fit parameters.
The results of the extended RTF model with the
rescaled $\sigma$ meson mass and coupling are given
by the open blue circles. Because this calculation cannot describe
the extended neutron density distribution at large radii sufficiently well,
the neutron skin thicknesses are systematically smaller than the
mean-field Hartree results with the same isovector interaction. 
However, the general trend is the same. Including the
$\alpha$-particle correlation 
leads to a further reduction of the neutron skin thickness in
the order of $0.02$~fm, which can be a substantial fraction of the
total neutron skin thickness. Thus, the correlation between $r_{\rm
  skin}$ and $L$ is modified when $\alpha$ cluster formation is taken
into account.

In conclusion, it was shown that an extended RTF model with explicit 
$\alpha$-cluster degrees of freedom predicts an appearance of
$\alpha$-particles on the surface of heavy nuclei and a reduction of the
neutron skin thickness depending on the neutron excess of the
nucleus. This behavior affects the $r_{\rm
  skin}$-$L$ correlation observed in conventional mean-field models. 
Therefore, the extraction of the parameter $L$ from measuring $r_{\rm skin}$ needs some
caution and the clusterization effect increases the systematic error.
Obviously, for more precise quantitative results on the
amount of $\alpha$-clustering on the nuclear surface and the
change of the neutron skin thickness due to $\alpha$-particle
correlations, improved calculations beyond the
extended RTF approximation, which also take pairing and
shell effects into account, have to
be performed in the future. The systematic variation of 
$\alpha$-particles abundancies on the nuclear surface should be studied
experimentally, e.g., by quasi-free ($p$,$p \alpha$) reactions \cite{Aum14}.

\begin{acknowledgments}
This work was supported by the Helmholtz Association (HGF) through the
Nuclear Astrophysics Virtual Institute (VH-VI-417). The author
gratefully acknowledges discussions with
T. Aumann, G. R\"{o}pke, T. Uesaka, and H.H. Wolter.
\end{acknowledgments}

\bibliography{nskin}

\end{document}